\documentclass[conference]{IEEEtran}
\IEEEoverridecommandlockouts
\usepackage{cite}
\usepackage{amsmath,amssymb,amsfonts}
\usepackage{algorithmic}
\usepackage{graphicx}
\usepackage{textcomp}
\usepackage{xcolor}
\usepackage{CJKutf8}
\usepackage{comment}
\def\BibTeX{{\rm B\kern-.05em{\sc i\kern-.025em b}\kern-.08em
    T\kern-.1667em\lower.7ex\hbox{E}\kern-.125emX}}
\usepackage{float}
\usepackage{placeins}
\usepackage{varioref}
\usepackage{stfloats}
\usepackage{multirow}

\begin{document}
\title{A Closed-loop Sleep Modulation System with FPGA-Accelerated Deep Learning} 

\author{
\IEEEauthorblockN{Mingzhe Sun, Aaron Zhou, Naize Yang, Yaqian Xu, Yuhan Hou, and Xilin Liu}
\IEEEauthorblockA{
\textit{Department of Electrical and Computer Engineering, University of Toronto}, Toronto, ON, Canada M5S 2E4 \\
Email: xilinliu@ece.utoronto.ca}
}

\maketitle

\begin{abstract}
\color{black}
Closed-loop sleep modulation is an emerging research paradigm to treat sleep disorders and enhance sleep benefits. However, two major barriers hinder the widespread application of this research paradigm. First, subjects often need to be wire-connected to rack-mount instrumentation for data acquisition, which negatively affects sleep quality. Second, conventional real-time sleep stage classification algorithms give limited performance. In this work, we conquer these two limitations by developing a sleep modulation system that supports closed-loop operations on the device. Sleep stage classification is performed using a lightweight deep learning (DL) model accelerated by a low-power field-programmable gate array (FPGA) device. The DL model uses a single channel electroencephalogram (EEG) as input. Two convolutional neural networks (CNNs) are used to capture general and detailed features, and a bidirectional long-short-term memory (LSTM) network is used to capture time-variant sequence features. An 8-bit quantization is used to reduce the computational cost without compromising performance. The DL model has been validated using a public sleep database containing 81 subjects, achieving a state-of-the-art classification accuracy of 85.8\% and a F1-score of 79\%. The developed model has also shown the potential to be generalized to different channels and input data lengths. Closed-loop in-phase auditory stimulation has been demonstrated on the test bench. 

\end{abstract}

\section{Introduction}

Sleep plays a critical role in a vast array of physiological and pathophysiological processes, including in neurodegenerative diseases such as Huntington’s and Alzheimer’s disease \cite{peter2015sleep}. New therapies will emerge through an improved understanding of sleep mechanisms. Sleep is composed of an alternation of rapid eye movement (REM) sleep and three non-REM (NREM) sleep stages (i.e., N1-3). Sleep stages can be classified using surface electroencephalogram (EEG). Causal investigations of sleep increasingly rely on closed-loop paradigms in which a real-time sleep stage classifier is used to deliver stage-specific stimulation \cite{liu2021system,liu2020fully,liu2016design,liu2015pennbmbi}. For example, studies have shown that auditory stimulation can be applied in phase with the prominent slow wave activity of NREM to enhance long-term memory \cite{ketz2018closed,ngo2013auditory}.

However, closed-loop sleep investigations are only feasible if the sleep stages can be detected accurately in real time and the stimulus signal can be delivered in phase with the sleep oscillation. The system should also minimize adverse effects on the subjects' sleep process, which excludes the use of rack-mounted instrumentation. These requirements motivate the development of a miniature system that supports closed-loop on-device sleep modulation and can be worn comfortably during sleep. 

\begin{figure}[!ht]
    \centering
    \includegraphics[width=0.45\textwidth]{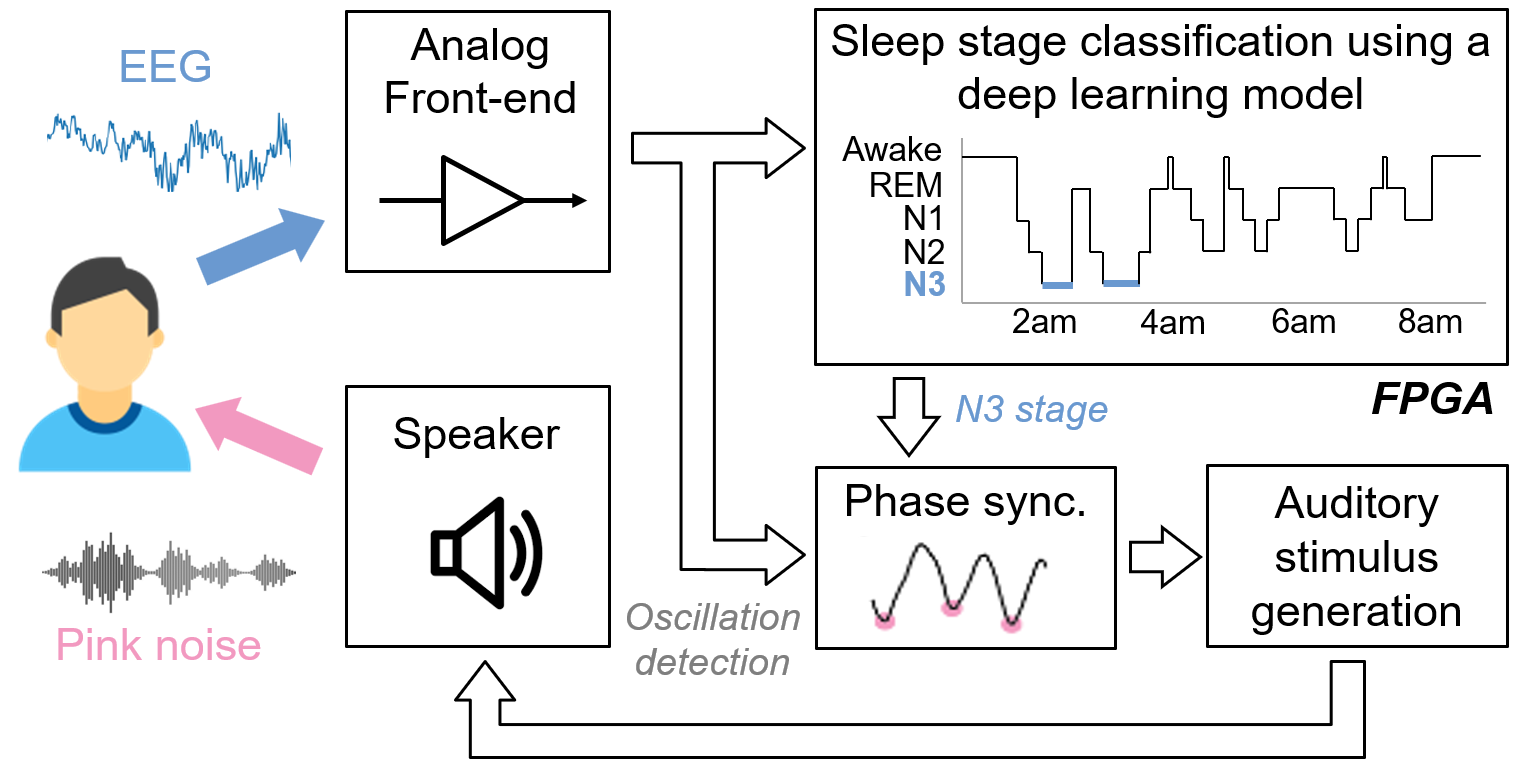}
    \vspace{-2mm}
    \caption{Block diagram of the proposed closed-loop sleep modulation system.}\vspace{-6mm}
    \label{fig:intro}
\end{figure}

In this work, we fill this important research gap by developing a closed-loop sleep modulation system that is self-contained and can be miniaturized. Fig. \ref{fig:intro} shows the high-level block diagram of the system and its operational principle. 
The system uses a single-channel EEG as input. 
Sleep stage classification is performed using a novel light-weight deep learning (DL) model implemented in a low-power field-programmable gate array (FPGA). Auditory stimulation is activated on the basis of the specific sleep stage and the detected sleep oscillation. 


The FPGA-accelerated DL model is a key component of the system. Although machine learning (including DL) algorithms have been developed to classify sleep stages, existing models have several common limitations: (1) demanding too much computational resources that are not available in energy-constrained wearable devices \cite{liu2021edge}; (2) using too many input channels that result in high power dissipation for signal acquisition and causing inconvenience in electrode placement; (3) using long time series as input (for example, more than a minute \cite{liu2021system}), which causes latency in real-time operation and thus is not suitable for closed-loop modulation. In this work, we develop a DL model that uses only one EEG channel as input with a segment of 20 or 30 seconds. A sliding window with overlap is used to further reduce inference latency. The overall model consists of only 1.28 M parameters, which makes it suitable for FPGA implementation.



\begin{figure*}[!t]
    \centering
    \includegraphics[width=0.95\textwidth]{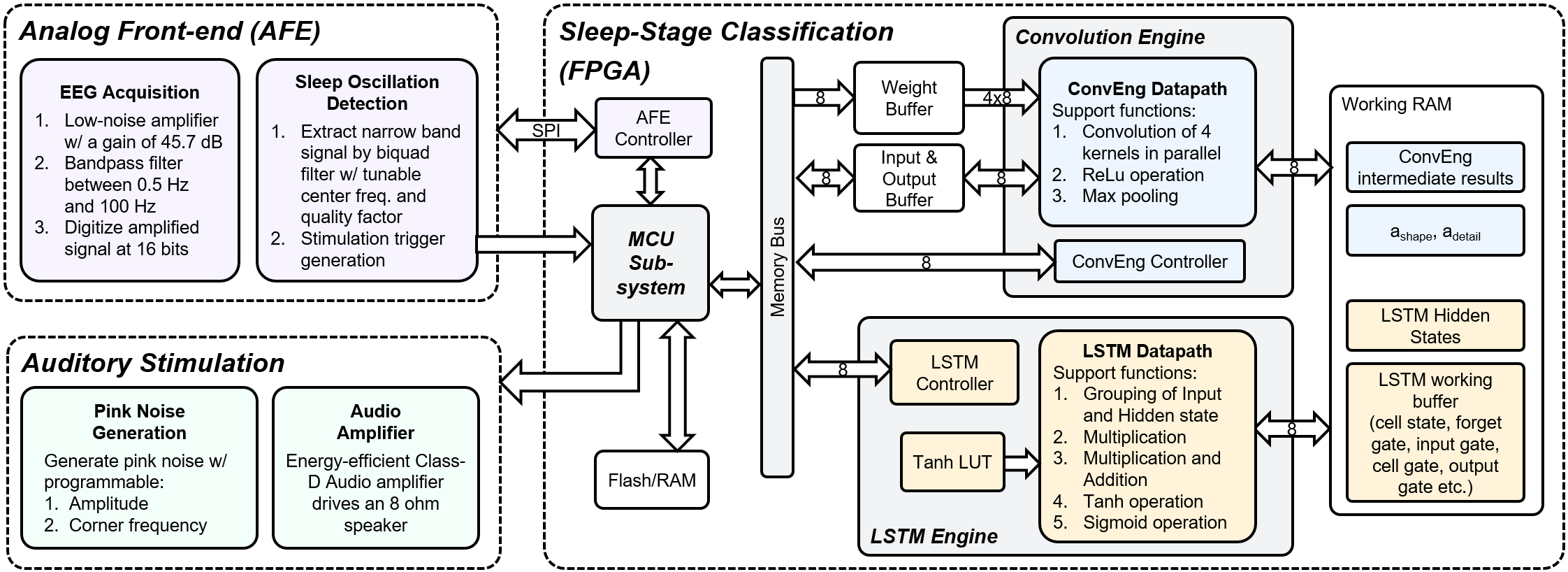}
    \vspace{-3mm}
    \caption{Top-level system architecture with key building blocks highlighted: (1) an AFE module for EEG signal acquisition and sleep oscillation detection; (2) a FPGA based module for sleep stage classification; and (3) an auditory stimulation generation module}\vspace{-3mm}
    \label{fig:sys_architecture}
\end{figure*}

The block diagram of the proposed sleep modulation system is illustrated in Fig. \ref{fig:sys_architecture}. The key building modules include: an analog front-end (AFE) module for EEG acquisition and oscillation detection, an FPGA module for DL based sleep stage classification and closed-loop control, a stimulator module for delivering auditory feedback, and peripheral modules such as Flash memory and power management units. 

The rest of the paper is organized as follows. Section III presents the DL model for the sleep stage classification. Section IV discusses the FPGA implementation of the DL model and the design of analog modules. Section V shows the experimental results. Finally, Section VI concludes the paper.

\section{Deep Learning Model for Sleep Stage Classification}

Machine learning models have been developed to classify sleep stages \cite{hussain2022quantitative,al2014methods,geethanjali2012time,diykh2016eeg}. Conventional machine learning models reply upon handcrafted features, including features in the frequency domain (e.g., fast Fourier transform \cite{hussain2022quantitative}), the time domain (e.g., change in slope sign \cite{geethanjali2012time}, waveform length \cite{diykh2016eeg}), or the time-frequency domain (e.g., discrete wavelet transform \cite{al2014methods}). Although machine learning models with hand-crafted features have proven their ability to automate sleep scoring, they often render poor generality when applied to different subjects and electrode placements. 
Recently, DL models have shown promising results in classifying sleep stages without hand-crafted feature selection \cite{movahedi2017deep,sors2018convolutional}. Deep belief networks \cite{movahedi2017deep} and convolutional neural network (CNN) \cite{sors2018convolutional} models have been developed using time-domain signals directly as input. These models have shown the ability to extract time-invariant features, but miss time-variant features. To capture time-variant features, such as sleep stage transitions, recurrent neural networks (RNNs) can be used \cite{seo2013obesity}.

\begin{figure}[!ht]
    \centering
    \includegraphics[width=0.42\textwidth]{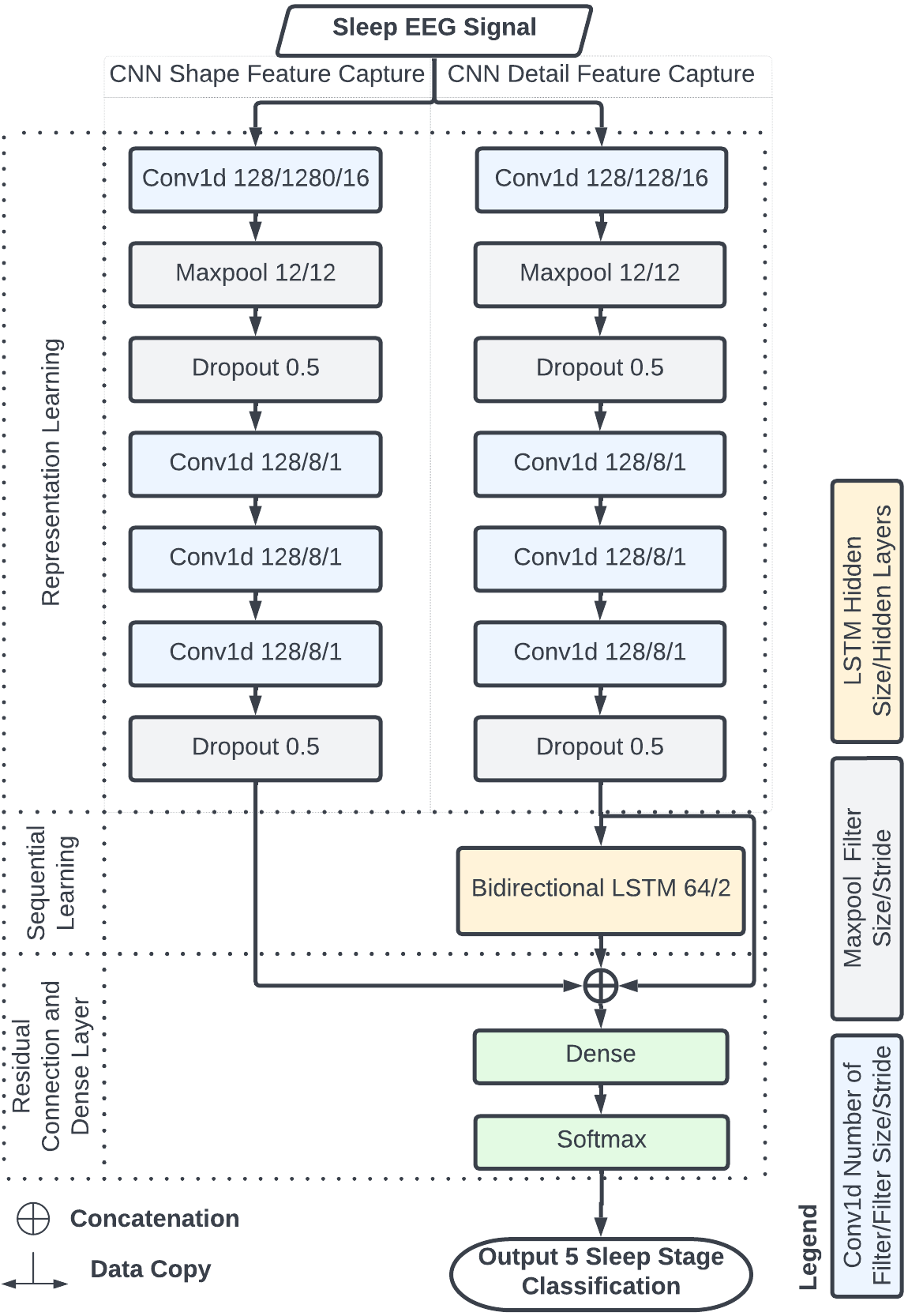}\vspace{-2mm}
    \caption{Architecture of the proposed DL model.}\vspace{-7mm}
    \label{fig:DL_model}
\end{figure}

In this work, we developed a hybrid DL model that takes advantage of both CNN and RNN. Fig. \ref{fig:DL_model} shows the model architecture. The model consists of three parts. The first part uses representation learning to capture time-invariant information from the input vector. The second part uses sequential learning to capture the sleep stage transition using features encoded in the first part. The third part consists of a dense network with residual connection to generate the prediction.  

{\color{black}
The representation learning part consists of two CNN paths, which are trained to learn features with different time scales. One path has a large filter size to capture general shape characteristics ($a_{shape}$) with low-frequency content, and the other path has a small filter size to capture detail shape characteristics ($a_{detail}$) with high-frequency content. Both CNN paths consist of four 1-D convolutional layers, two dropout layers, and one max-pooling layer. Each 1-D convolutional layer is followed by batch normalization and a rectified linear unit (ReLU) activation function. 
Two dropout layers are added to reduce overfitting. To demonstrate the generalizability of the model, we designed the representation learning part to accommodate input EEG signals with a length of 20-sec or 30-sec. This is made possible by adjusting the max-pooling layer and the dropout layers in the two CNN paths, providing a similar output data length for the sequential learning part. 
}



{\color{black}
Transitions between sleep stages often occur in patterns. 
Therefore, we designed a sequential learning part using a bidirectional LSTM network to capture the sleep transition. Only $a_{detail}$ is used as input to the LSTM model for sequential learning, instead of concatenating $a_{detail}$ and $a_{shape}$. This allows us to obtain optimal performance at a low computational cost. The sequential learning part outputs the final forward hidden state $h_{f}$ and the first reverse hidden state $h_{r}$ of the extracted features. 
}


{\color{black}
The last part of the model is a dense network that generates the final prediction. Both the outputs of the representational learning part ($a_{shape}$, $a_{detail}$) and the sequential learning part ($h_f$, $h_r$) are taken as input.  $a_{shape}$ and $a_{detail}$ are given to the dense network as residual connections. They add frequency content that is degraded in sequential learning. $h_f$ and $h_r$ are concatenated to provide sufficient time domain features in both the forward and reverse directions. 
}

{\color{black}
\section{Hardware Design}
\subsection{FPGA-Accelerated Deep Learning Model}
The DL model was implemented on Zynq®-7000 XC7Z020-CLG484-1 from AMD Xilinx. The blocks of the FPGA implementation are depicted in Fig. \ref{fig:sys_architecture}. A microcontroller unit (MCU) block integrated in the FPGA manages system control. The weights of the DL model are stored in the Flash memory and loaded to the data engine for processing. The MCU monitors the status of the data engine and loads the corresponding weights into the buffers. 

The convolution engine processes the EGG data from the MCU subsystem and generates $a_{detail}$ and $a_{shape}$. Subsequently, the LSTM engine calculates $h_{f}$ and $h_{r}$ from $a_{detail}$. Then the convolution engine performs the dense operation with the above results and sends the result to the MCU through the output buffer. Finally, the MCU performs softmax and obtains the final detection result. To minimize memory access, the convolution engine shown in Fig. 2 processes 4 kernels in parallel, with ReLu and max pooling operations. The address generator in the controller enables flexible data arrangement of input and output data, so that no additional data moving or reordering is required.  

}
The LSTM engine concatenates the input and hidden states of each layer, so that the convolution engine is used to generate intermediate results of the forget gate, input gate, cell gate and output gate in scratch memory in working RAM. Then the LSTM datapath performs Sigmoid and Tanh operations, as well as the following multiplications and additions, to generate the updated cell state and hidden state. An optimized interpolation algorithm with a 38-enity Tanh lookup table is used to implement Tanh and sigmoid operations. The interpolation algorithm shares the same multiplier with other LSTM operations for lower hardware complexity. 

For simplicity of the memory system and better performance, memory blocks are designed with simple dual-port memory. The engine could process 20-sec input data within 1sec when running at 20 MHz clock. It provides flexibility to support multiple EEG inputs as well as future algorithm enhancement. {\color{black}To further reduce the hardware cost, we applied static quantization for both weight and activation to signed 8 bit. We bench-marked three calibration methods, MinMax, entropy, and percentile. Appropriate data shifting and saturation operations are performed in the operations. Table \ref{table_parameters} summarizes the final resources used for the FPGA implementation.}

\begin{table}[!ht]
\begin{center}\vspace{-2mm}
\caption{Resources for FPGA Implementation}\label{table_parameters}\vspace{-1mm}
\begin{tabular}{|l|l|l|l|}
\hline
                  & \textbf{Parameters} & \textbf{\begin{tabular}[c]{@{}l@{}}Multiplication\\ Operation\end{tabular}} & \textbf{\begin{tabular}[c]{@{}l@{}}Memory\\ Required\end{tabular}} \\ \hline
CNN-shape         & 425984              & 55836672                                                                    & 7936                                                               \\ \hline
CNN-detail        & 278528              & 9830400                                                                     & 7936                                                               \\ \hline
LSTM              & 196608              & 2960640                                                                     & 1920                                                               \\ \hline
Residual \& dense & 376832              & 376832                                                                      & -                                                                  \\ \hline
\textbf{Total}    & \textbf{1277952}    & \textbf{69004544}                                                           & \textbf{17792}                                                     \\ \hline
\end{tabular}\label{table_parameters}\vspace{-2mm}
\end{center}
\end{table}

{\color{black}
\subsection{Analog Front-end Module Design}
The AFE consists of an EEG acquisition path and a sleep oscillation detection path. The EEG acquisition path uses a commercial neural amplifier (RHD2216, Intan Technologies) for signal amplification and digitization. The amplifier has a gain of 49.5 dB and a digitization resolution of 16 bits. The sleep oscillation detection path uses a 4th-order biquad filter, as shown in Fig. \ref{fig:ckt}(a). 
\begin{figure}[!ht]
\vspace{-4mm}
    \centering
    \includegraphics[width=0.4\textwidth]{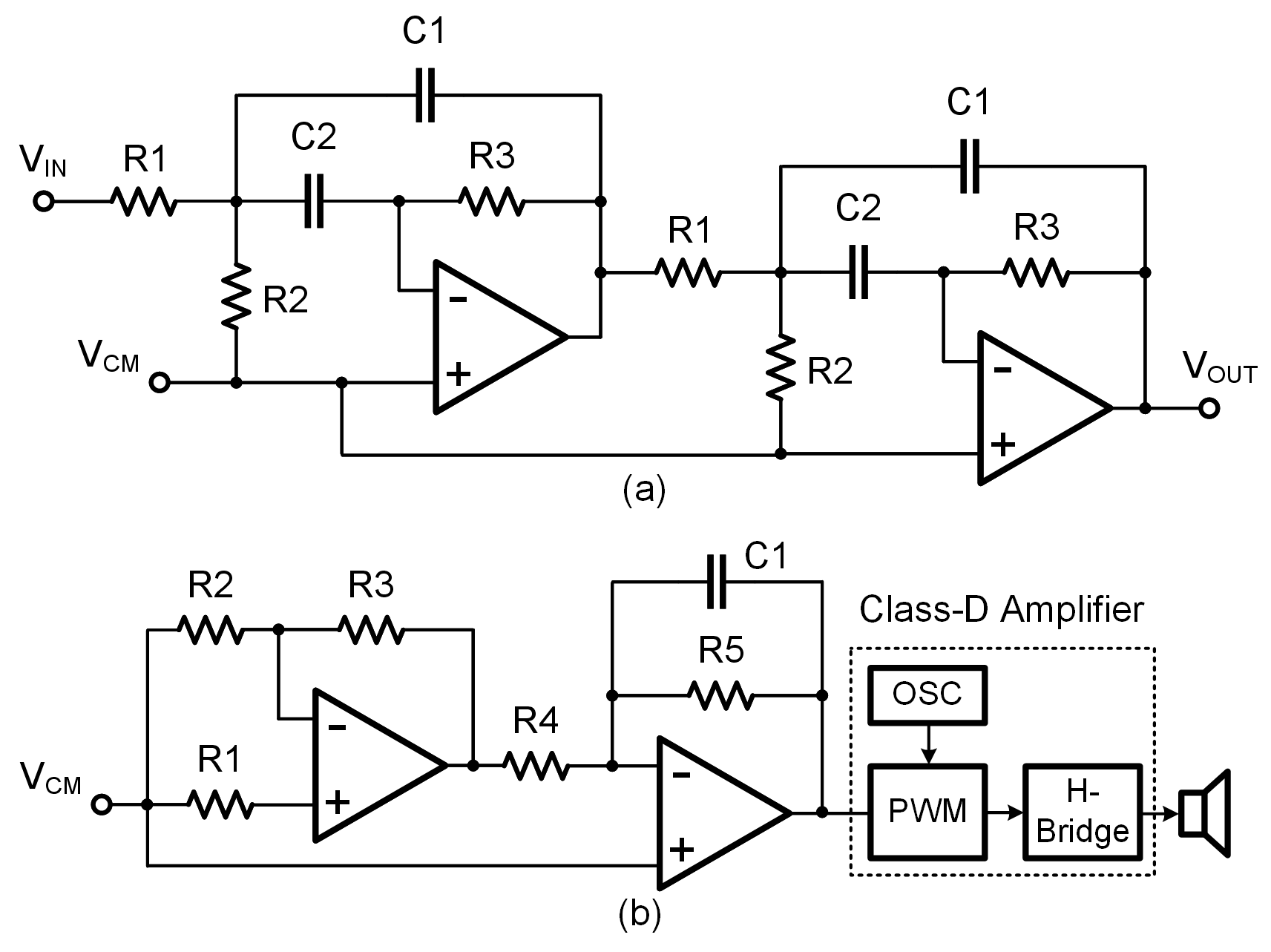}
    \vspace{-2mm}
    \caption{Simplified circuit schematic of the developed (a) 4th-order Biquad filter for slow oscillation detection and (b) pink noise generation. }
    \label{fig:ckt}
\end{figure}
The circuit uses only one operational amplifier per biquad core to save power consumption \cite{sedra2004microelectronic,liu2015design}. The transfer function of the biquad filter is given by:
\begin{equation}
    H(s) = \frac{-s(\frac{\alpha}{C_1R_{eq}})}{s^2 + s(\frac{1}{C_1} + \frac{1}{C_2})\frac{1}{R_3} + \frac{1}{C_1C_2 R_3R_{eq}}}
\end{equation}
where $R_{eq} = (1/R_1+1/R_2)^{-1}$ and $\alpha =R_4/R_1$. The center frequency is 
\begin{equation}
\omega_0 = 1/\sqrt{C_1C_2R_3R_{eq}}
\end{equation}
and the quality factor is given by
\begin{equation}
    Q = [\frac{\sqrt{C_1C_2R_3R_{eq}}}{R_3}(\frac{1}{C_1}+\frac{1}{C_2})]^{-1}
\end{equation}

A comparator is used to detect the zero-crossing point of the filtered oscillation signal. The detection signal is sent to the FPGA and a programmable delay is added before triggering the stimulation. 
}

\subsection{Auditory Stimulation Module Design}
The pink noise generation was implemented in the analog domain, as shown in Fig. \ref{fig:ckt}(b). A 150 k$\Omega$ resistor is used as the source of white noise, which is amplified and filtered by a first-order low-pass filter to generate pink noise. The frequency characteristics can be further shaped by the filter. An energy-efficient class-D amplifier (TPA2005D1, Texas Instruments) was used to drive an 8 $\Omega$ piezo transducer speaker (AS01008MR-2-R, PUI Audio).

\section{Experimental Results}
\subsection{Validation of the DL Model}

A public database from the Montreal Archives of Sleep Studies (MASS)\cite{o2014montreal} was used to train and test the DL model. The MASS database contains 5 subsets (SS1-5) of adult polysomnography recordings, which were labeled by experts. We evaluated our model on the subset SS2 \cite{ceams2022sstwo} and SS3 \cite{ceams2022ssthree}. SS2 contains recordings of 19 subjects labeled per 20-sec EEG epoch based on the Rechtschaffen and Kales (R\&K) standard; SS3 contains recordings of 62 subjects labeled per 30-sec EEG epoch based on the AASM standard. All EEG recordings have a sampling rate of 256Hz. Our model was developed to predict sleep stages based on the AASM standard. To convert the labels from R\&K to AASM, N-REM stages 3 and 4 are combined to N3 and unknown stages are not taken into account. 



{\color{black}
To evaluate the model performance, we adopted a leave-one-subject out cross-validation strategy for SS2, and leave-two-subjects-out for SS3 since it contains more data. 10\% of the test subjects' data were used for fine-tuning per validation, and the remaining 90\% of the data were used for testing. All test data were excluded from training. Adam optimizer was used for training with $lr = 10^{-4}$, $beta1 = 0.9$, $beta2 = 0.999$ for 100 epochs. L2 weight decay was adopted to prevent overfitting with a value of $10^{-3}$. We used a batch size of 256 for general training. The sequence length of 3 is used in sequential learning. The extracted features of the previous two segments and the current segment were used as input to the sequential learning section. We used overall accuracy (ACC), macro F1-score, Cohen's Kappa coefficient ($k$) and per-class accuracy to evaluate the performance of our model. 

}



Table \ref{table_quant} summarizes the performance of our model based on the evaluation in MASS-SS2. 
\begin{table}[!ht]
\centering \vspace{-3mm}
\caption{Performance Before and After Quantization} \label{table_quant} \vspace{-3mm}
\begin{center}
\begin{tabular}{|l|ccc|ccccc|}
\hline
\centering
\multirow{1}{*}{Quan-}  & \multicolumn{3}{c|}{Overall Metrics} & \multicolumn{5}{c|}{Per-class ACC} \\ 
tization & ACC        & F1         & k          & W     & N1    & N2   & N3   & REM  \\ \hline
 Before & 85.5       &   76.6         & 78.6       &  83.2     & 43.5      & 89.3     &  84.2    &  82.7    \\  
 After & 84.9       & 75.8       & 77.8       & 82.3 & 42.0 & 89.1 & 83.9 & 81.8 \\\hline
\end{tabular}
\end{center}\vspace{-3mm}
\end{table}
Performance before and after 8-bit quantization is shown. The result suggests a marginal degradation of approximately 1\% for significant hardware savings. Table \ref{table_comparison} shows the performance of our model evaluated in MASS-SS2 and SS3 compared to the state-of-the-art works. On average of the 81 subjects, the accuracy of our model is 85.8\% and the F1-score is 79\%, which are comparable to the state-of-the-art DL models without hardware implementation.  


\begin{table}[!ht]
\centering\scriptsize
\renewcommand{\arraystretch}{1.2}
\caption{Comparison with state-of-the-art models}\label{table_comparison}
\begin{tabular}{|c|c|c|c|ccc|}
\hline
\multirow{2}{*}{Data}                                            & \multirow{2}{*}{Year} & \multirow{2}{*}{Publication} & \multicolumn{1}{l|}{\multirow{2}{*}{Methods}} & \multicolumn{3}{c|}{Overall Metrics}                         \\ \cline{5-7} 
                                                                     &                       &                              & \multicolumn{1}{l|}{}                         & \multicolumn{1}{c|}{ACC}  & \multicolumn{1}{c|}{F1}   & k    \\ \hline
\multirow{5}{*}{\begin{tabular}[c]{@{}c@{}}MASS\\ -SS3 \\ (30-sec) \end{tabular}} & 2021                  & EOGNET\cite{fan2021eognet}                       & CNN+RNN                                       & \multicolumn{1}{c|}{83.1} & \multicolumn{1}{c|}{76.4} & 75 \\ \cline{2-7} 
                                                                     & 2020                  & IITNet\cite{seo2020intra}                      & CNN+RNN                                       & \multicolumn{1}{c|}{86.6} & \multicolumn{1}{c|}{80.8} & 80   \\ \cline{2-7} 
                                                                     & 2017                  & DeepSleepNet\cite{supratak2017deepsleepnet}                 & CNN+RNN                                       & \multicolumn{1}{c|}{86.2} & \multicolumn{1}{c|}{81.7} & 80   \\ \cline{2-7} 
                                                                     & 2020                  & TinySleepNet\cite{supratak2020tinysleepnet}                 & CNN+RNN                                       & \multicolumn{1}{c|}{87.5} & \multicolumn{1}{c|}{83.2} & 82   \\ \cline{2-7} 
                                                                     &                       & This work                    &         CNN+RNN                                       & \multicolumn{1}{c|}{86.1}     & \multicolumn{1}{c|}{80.0}     &     79.4 \\ \hline
\multirow{3}{*}{\begin{tabular}[c]{@{}c@{}}MASS\\ -SS2 \\ (20-sec)\end{tabular}} & 2021                  & MetaSleepLearner\cite{banluesombatkul2020metasleeplearner}             & CNN                                           & \multicolumn{1}{c|}{77.3} & \multicolumn{1}{c|}{69.9} & 68   \\ \cline{2-7} 
                                                                     & 2020                  & TinySleepNet\cite{supratak2020tinysleepnet}                 & CNN+RNN                                       & \multicolumn{1}{c|}{82.6} & \multicolumn{1}{c|}{75.5} & 75   \\ \cline{2-7} 
                                                                     &                       & This work                    &         CNN+RNN                                       & \multicolumn{1}{c|}{84.9} & \multicolumn{1}{c|}{75.8} & 77.8   \\ \hline
\end{tabular}
\end{table}

\subsection{Validation of Analog Modules and Closed-loop Operation}
\begin{figure}[!ht]
    \centering\vspace{-3mm}
    \includegraphics[width=0.48\textwidth]{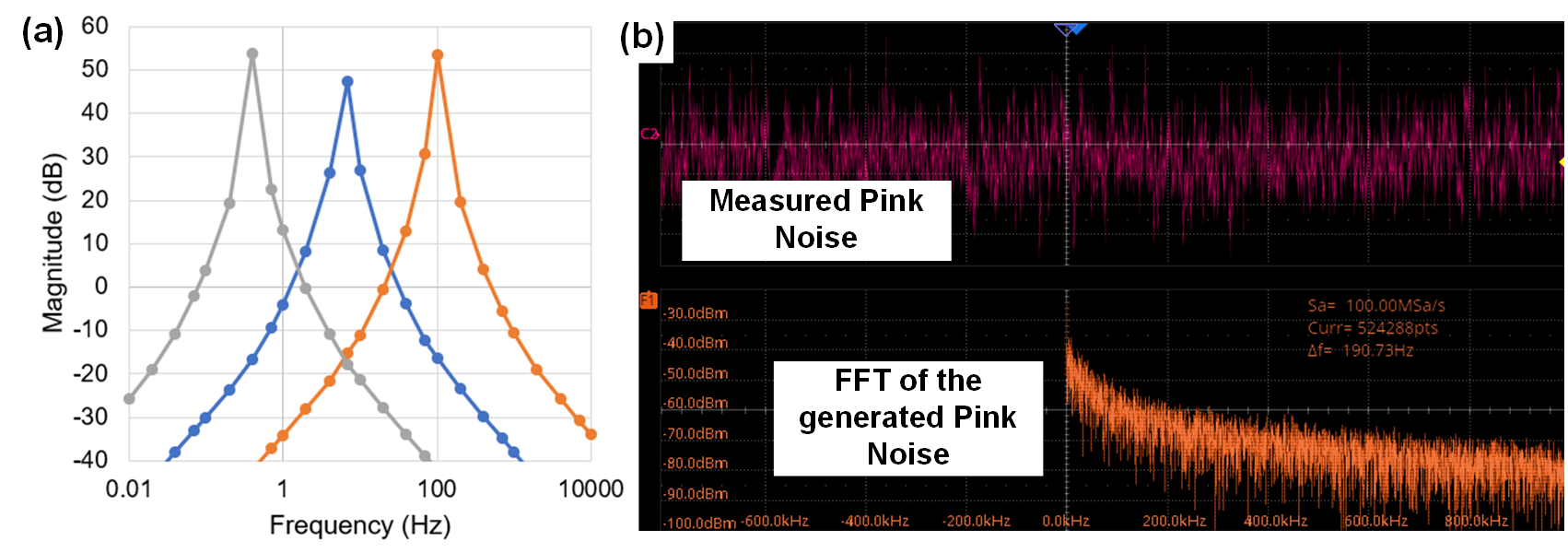}
    \vspace{-2mm}
    \caption{(a) Experimentally measured frequency response of the biquad filter. (b) Time-domain and spectrum of the generated pink noise.}
    \label{fig:exp_2}
\end{figure}

\begin{figure}[!ht]
    \centering
    \includegraphics[width=0.48\textwidth]{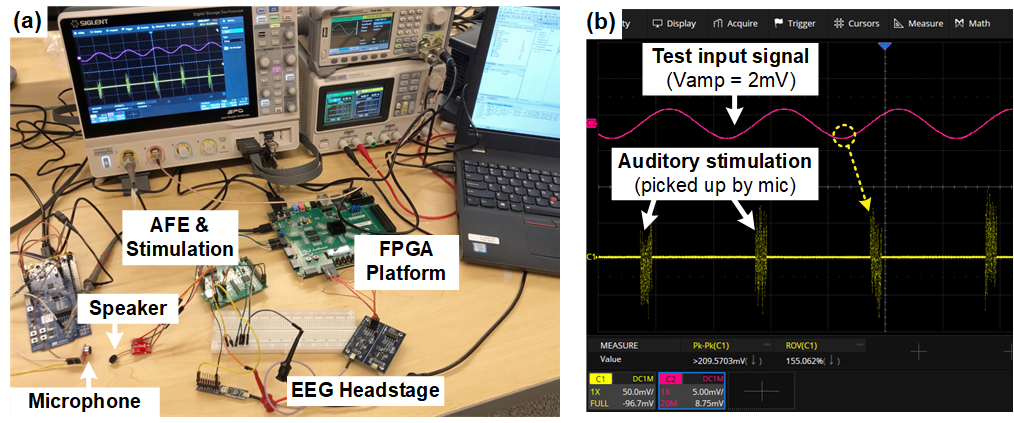}
    \vspace{-2mm}
    \caption{(a) Experimental setup for validating of closed-loop auditory stimulation. (b) Oscilloscope screenshot showing 1 Hz test input signal along with triggered auditory stimulation picked by microphone.}\vspace{-3mm}
    \label{fig:exp_1}
\end{figure}

The analog front-end and auditory stimulation modules have been fully characterized on a bench. Fig. \ref{fig:exp_2} shows the experimentally measured results for biquad filtering and pink noise generation. 
Closed-loop auditory stimulation has also been demonstrated (at the moment without the DL model). Fig. \ref{fig:exp_1} illustrates the experiment, in which the system successfully detected the sinusoid test signal of 1 Hz and triggered synchronized auditory stimulation (picked up by a microphone). 

\section{\color{black}Conclusion}
This paper presents a first-of-its-kind closed-loop auditory sleep modulation system, featuring a FPGA-accelerated DL model that delivers real-time sleep stage classification with state-of-the-art performance. This design holds great promise in enabling novel sleep research paradigms with potential for clinical translation. 

\bibliographystyle{IEEEtran}
\bibliography{ref}

\end{document}